\documentstyle[12pt]{cospar}
\def\title#1{\relax\vspace*{2cm}{\large{\bf #1}}\par\vspace*{13.5pt}}
\def\author#1{{#1}\par\vspace*{13.5pt}}
\def\affil#1{{\it #1}\par}
\def\abstract{\vspace*{27pt}ABSTRACT\par\relax}
\def\section#1{\par{#1}\par}
\def\subsection#1{\par\underline{#1}\par}
\def\subsubsection#1{\par\underline{#1.}\ \ }
\def\acknow{\par ACKNOWLEDGMENTS\par}
\newenvironment{references}{\section{REFERENCES}\vspace*{.5cm}%
 \parindent=0pt\frenchspacing%
 \parskip=1pt plus 1pt minus 1pt%
 \interlinepenalty=1000\tolerance=400%
 \pretolerance=10000\hyphenpenalty=10000%
 \everypar={\hangindent=1.6pc}
}{}

\begin{document}
\input{psfig.tex}

\title{MULTIWAVELENGTH PROPERTIES OF BLAZARS}

\author{C. Megan Urry$^1$}
\affil{$^1$Space Telescope Science Institute,
3700 San Martin Drive, Baltimore MD 21218, USA}

\abstract

The multiwavelength spectra of blazars appear to be dominated by nonthermal 
emission from a relativistic jet oriented close to the line of sight. The
recent detection of many blazars at gamma-ray energies strongly supports
this scenario. High quality multiwavelength monitoring data for the brightest
one or two blazars suggest the optical through X-ray continuum is synchrotron 
emission from an inhomogeneous jet. The gamma-rays are likely due to Compton
scattering of lower energy photons, either from within the jet or from the
surrounding gas. The physical properties of the jet and the way in which it
is produced are still largely a mystery but are probably related in some way 
to accretion onto a central supermassive black hole. There is little direct 
observational
evidence for accretion disks in blazars, although there is evidence for
winds which might emanate from disks.

\section{BLAZARS IN RELATION TO OTHER AGN}

Blazars are a special kind of active galaxy characterized by very
rapid variability, high and variable polarization, superluminal
motion, and very high luminosities --- in short they are the most
``active'' kind of AGN.
Given that the theme of this COSPAR session is accretion disk phenomenology,
I begin by describing the relation of blazars, which in general do not 
show direct observational evidence of accretion disks, to AGN that do.

\footnotesize
\begin{figure}
\centerline{\psfig{figure=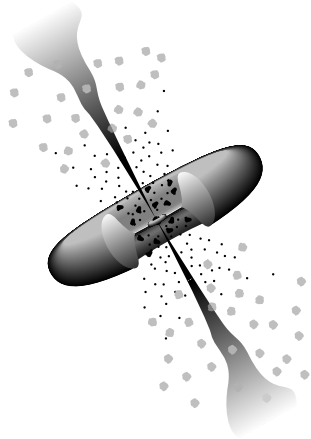,height=2.32in}}


\caption{
Schematic drawing of the current paradigm for AGN (not
to scale; example lengths in parentheses). 
Surrounding the central black hole 
(for $M=10^8 M_\odot$ black hole,
$R_S\equiv 2 GM /c^2 \sim3\times10^{13}$~cm)
is a luminous accretion disk
($\sim 1 - 30 \times 10^{14}$~cm).
Broad emission lines are produced in clouds orbiting above the disk 
(at $\sim2 - 20 \times 10^{16}$~cm) and perhaps by the disk itself. 
A thick dusty torus (inner radius $\sim10^{17}$~cm;
or warped disk) obscures the
broad-line region (BLR) from transverse lines of sight; some continuum and
broad-line emission can be scattered into those lines of sight by hot
electrons that pervade the region. A hot corona above the accretion disk may
also play a role in producing the hard X-ray continuum. Narrow lines are
produced in clouds much farther from the central source
($10^{18} - 10^{20}$~cm).
In radio-loud AGN,
radio jets emanate from the region near the black hole
(extending from $10^{17}$ to several times $10^{24}$~cm, 
a factor of ten larger than the largest galaxies),
initially at relativistic speeds. (From Urry \& Padovani 1995;
copyright Astronomical Society of the Pacific, reproduced with permission.)
\label{fig:cartoon}
}
\end{figure}
\normalsize

Figure~1 illustrates the current AGN paradigm, in which a supermassive
black hole provides the central power source.
An equatorial accretion disk provides a means for funneling
matter onto the black hole. The surrounding emission line clouds, both
broad emission line clouds and narrow emission line clouds 
farther out, are less apparent in the spectra of blazars than 
in other AGN but both narrow and broad lines are sometimes seen.
An optically thick screen, shown in Figure~1 as a thick torus, obscures
the central continuum and broad-line emission along some lines of sight.
Some fraction of AGN, perhaps 10\%, are radio loud, 
and these have jets which are at least initially relativistic.

There are strong indications that blazars
are the radio-loud AGN seen more or less end-on down the jet,
so that the bulk relativistic motion of the emitting plasma causes
radiation to be beamed in a forward direction, making the 
variability appear more rapid and the luminosity appear higher than in
the rest frame (Urry \& Padovani 1995, and references therein).
First, virtually every blazar exhibits superluminal motion in 
high-resolution radio maps (Vermeulen \& Cohen 1994), 
which is easily explained by relativistic
bulk motion along the line-of sight. Second, they are all highly
polarized, at least some of the time. Indeed, the classical definition
of blazars as optically violently variable AGN turns out to be consistent
with defining them by high polarization (excluding radio-quiet AGN 
polarized by scattering), or by superluminal motion, 
or most recently, by their strong gamma-ray emission (taking duty cycle
into account).
Third, blazars are well matched with radio galaxies, which are known
to have jets more nearly in the plane of the sky,
in terms of number and luminosities, as well as in terms of 
properties not affected by beaming (host galaxy, environment, and so on).
Their radio variability, multiwavelength variability, and polarization
characteristics can all be well explained by shocks in an aligned
relativistic jet.

Blazars have very high brightness temperatures, based on the rapid
variability of their radio and optical emission. This is described
in an excellent review article (Wagner \& Witzel 1995) so I say 
nothing more here about intra-day variability;
the remainder of this talk concerns multiwavelength properties of 
blazars, particularly at the higher energies.

\section{GAMMA-RAY EVIDENCE FOR BULK RELATIVISTIC MOTION}

The extreme characteristics of blazars can be accounted for by the
unification hypothesis --- that is, by saying that blazars are 
radio galaxies with jets pointing at us rather than in the plane of
the sky. The confirmation of this hypothesis was
the discovery with the EGRET experiment on the Compton Gamma-Ray Observatory
that blazars are very strong gamma-ray emitters (Fichtel 1994,
von Montigny et al. 1995). 

The superluminal quasar 3C~279, the first gamma-ray blazar 
discovered with EGRET, was at the time of its discovery one of the brightest 
gamma-ray sources in the sky (Hartman et al. 1992). At that epoch, 1991 June, 
it was undergoing a flare, more than doubling its gamma-ray intensity 
(at $\sim1$~GeV) in a few days (Kniffen et al. 1993). 
The Parkes radio source 1622-297 has flared even more dramatically, becoming 
5-10 times brighter than the high state of 3C~279 in 1991 June, 
doubling or quadrupling its flux in half a day (Mattox et al. 1997). 
Other blazars have shown similar rapid flaring at gamma-ray energies; 
basically, any blazar whose count rate is high enough to detect on 
short time-scales with EGRET is seen to vary on those time-scales.
Doubling times as short as minutes have been seen at TeV energies in 
one blazar, Mrk~421 (Gaidos et al. 1996).

That so many high energy photons emerge from what appears to be such a compact
volume leads to the conclusion that the gamma rays are relativistically 
beamed. The argument goes like this: 
the emission region cannot be as compact as it appears
(i.e., the gamma-ray photon density cannot be as high) or
the gamma-rays would interact with ambient X-ray photons to make pairs,
thus preventing the observed gamma-ray emission. With beaming, the 
rest-frame gamma-ray photon densities are dramatically smaller than
implied if the luminosity is emitted isotropically, and the source dimension
is also somewhat larger. Thus, that gamma-rays are observed at all
implies that blazars are beamed (Dondi \& Ghisellini 1995).

\section{THE SIGNIFICANCE OF BLAZARS FOR UNDERSTANDING AGN AND BLACK HOLES}

If blazars are such a special orientation of AGN, they are clearly a minority
event, and an accidental one at that. So understanding blazars for the sake
of blazars alone is not especially important. Because of their special
orientation, however, blazars offer a direct probe of the unknown
and very interesting physics of energy production in the centers of AGN,
probably associated with the presence of a supermassive black hole.
In this sense, understanding blazar emission may be more fundamental --- 
more directly related to the physics of the central black hole ---
than the distribution of broad-emission-line clouds or even the
thermal radiation from an accretion disk.

The ultimate goal of blazar research is to learn how
energy is extracted from the black hole. In a more immediate sense,
the goal is to understand how relativistic jets, which represent a 
tremendous amount of kinetic energy, are formed and accelerated.
We are still at a very early stage in this endeavor, trying at this point
simply to understand how much energy is involved and what the physical 
state of the jet is.
Specifically, we are trying to measure the energy densities of particles,
magnetic fields, and radiation, as well as the velocity field of the jet.
These must be inferred since the observed radiation depends strongly on
the amount of relativistic beaming, as well as the intrinsic properties
of the jet.

Were we able to resolve the jets at wavelengths from radio through gamma-rays,
this problem would be much simpler. But because the scales involved are
probably $10^{-8}$--$10^{-5}$~arcsec for regions emitting at or above
optical frequencies, current technology precludes direct imaging. Instead,
we infer the jet structure from variability as a function of wavelength.
Time-scales for the fastest (observed) variability range from minutes 
at the highest energies to days or weeks at optical or radio
wavelengths. This talk describes the results from recent multiwavelength 
monitoring campaigns on a few blazars, which are leading to important
progress in this area.

\section{WHAT IS NOT KNOWN ABOUT BLAZARS}

Two important scientific questions about blazars are partially 
but not completely answered at this point. 
First, how is the gamma-ray spectrum produced? In many cases 
the gamma-rays dominate the bolometric luminosity, at least during flare
states, so gamma-rays are fundamental to understanding 
how the bulk of the energy is produced in blazar jets.

Second, there are several sub-classes of blazars with distinct spectral
characteristics but the physical origin of this difference is not understood.
Some blazars have very strong emission lines and so are classified as
quasars while those with very weak emission lines are classified as 
BL~Lac objects, and among the BL Lac objects, there are two flavors 
(see below) with different continuum shapes.
The differences among these types of blazars is likely a
clue to their underlying structures.
 
\section{MULTIWAVELENGTH SPECTRA OF BLAZARS}

Substantial progress in understanding blazars has come from
multiwavelength spectral studies. Early single-epoch studies
led to the development of jet models for production of the broad-band
radio through X-ray continuum
(e.g., Marscher 1980, K\"onigl 1981, Ghisellini et al. 1985).
More recently, such models have been extended to GeV energies to 
explain the EGRET observations.

Figure~2 shows a schematic representation of the 
observed radio through ultra-high-energy gamma-ray continuum, in
power per logarithmic band ($\nu F_\nu$), for the three kinds of blazars. 
The low-frequency component is likely due to synchrotron radiation and
the high-frequency component to Compton-scattering of lower energy photons
by the same electron population (see discussion of 3C~279 below).

\footnotesize
\begin{figure}
\centerline{\psfig{figure=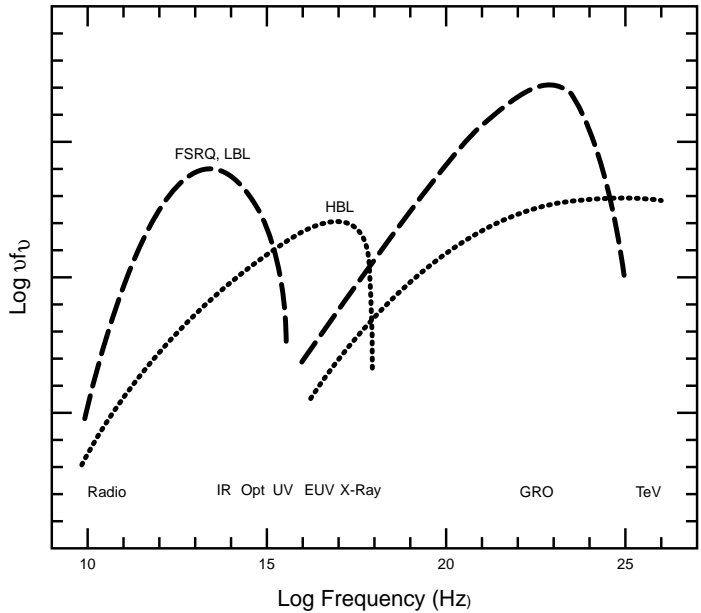,height=3.2in}}


\caption{
Spectral energy distributions for three kinds of blazars. The
synchrotron power of strong emission line blazars (FSRQ) and 
low-frequency peaked blazars (LBL) peaks at submillimeter to infrared 
wavelengths, while that of high-frequency peaked blazars (all known HBL are
BL Lac objects) peak at UV to X-ray wavelengths. The Compton powers
peak at GeV energies for FSRQ and LBL and at much higher (TeV) energies
for HBL. In general, FSRQ and LBL (dashed lines) are more luminous than HBL
(dotted lines), so that the wavelength of the peak power output 
correlates with luminosity.
}
\end{figure}
\normalsize

The weak-lined blazars, or BL Lac objects, fall into two categories,
defined by Padovani \& Giommi (1995) as ``Low-frequency peaked BL Lacs''
(LBL) and ``High-frequency peaked BL Lacs'' (HBL) depending on whether 
$\alpha_{\rm rx} \equiv \log (F_{\rm 5~GHz} / F_{\rm 1~keV})/ 7.68$
is greater than or less 
than 0.75, respectively. Most HBL have been found in X-ray surveys, and so
have been known previously as XBL (X-ray-selected BL Lac objects), while
most LBL have been found in radio surveys and so are also known as RBL. As such
surveys go deeper, however, the mix of types in a given survey will change,
with increasing numbers of LBL in X-ray surveys and more HBL in radio surveys,
hence the need for a quantitative and clear distinction between the two
types. At this point, it is not clear whether there exist two distinct
classes of BL Lac object or whether there is a continuous distribution of
spectral shapes between the classically discovered LBL and HBL.

The observed differences in continuum spectral
shape (Fig.~2), are that
the synchrotron power of LBL peaks at submm to IR 
wavelengths while that of HBL peaks at UV to X-ray wavelengths, and
the Compton components peak at GeV energies for LBL and at much 
higher (TeV) energies for HBL. 
HBL tend to be fainter EGRET sources than LBL even though they 
are a lower redshift population; their ratio of peak gamma-ray flux 
to peak synchrotron flux is around one or less. 

The strong-emission-line blazars are denoted by FSRQ 
(Flat-Spectrum Radio Quasars); the label FSRQ is more or less equivalent
to ``blazar'' since essentially all are highly variable and 
(at least some of the time)
highly polarized, as well as superluminal. 
The continuum shapes of FSRQ
are very similar to those of LBL (Sambruna et al. 1996), 
with synchrotron peaks at $10^{13}$--$10^{14}$~Hz and 
Compton peaks at $10^{22}$--$10^{23}$~Hz.

In general, FSRQ and LBL are more luminous than HBL, so that 
the wavelength of the peak power output increases with luminosity.
Also, for FSRQ and LBL the ratio of Compton to synchrotron power is higher
than for LBL, at least in the high state, so that the Compton power increases
proportionately more than the synchrotron power with increasing luminosity.
This is illustrated with real multiwavelength spectra in 
a paper by Sambruna et al. (1996), who discuss possible physical 
connections among FSRQ, LBL, and HBL.
One popular hypothesis is that LBL are viewed at smaller angles than
HBL, so that the difference is purely an orientation effect. Sambruna et al.
conclude to the contrary that there must instead be intrinsic differences
because for plausible emission models it is not possible to shift 
the wavelength of the peak emission by as much as four orders of magnitude.

\section{MULTIWAVELENGTH VARIABILITY OF BLAZARS}

The amplitude of variability in blazars is
greater at frequencies higher than the peak of the synchrotron 
spectrum, and probably this is the case in the Compton component as
well (although the data are more sparse). 
This explains in a phenomenological way why LBL are highly variable
in the optical while HBL are not: for LBL, optical emission lies above 
the peak frequency while for HBL it lies below.
At X-ray energies, on the other hand, HBL are among the
most rapidly variable AGN known. (Unfortunately, LBL are relatively faint
in the X-ray and so their variability has not been observed to equivalent
levels.)

This points out a significant selection effect, relevant to both variability
and spectroscopic studies currently available, 
due to the relative brightnesses at wavelengths
of interest. Most high-energy X-ray data are available for
HBL, while those blazars with well-studied radio and optical variability, 
particularly intraday variability, tend to be LBL. Also, the strongest
EGRET sources are LBL and the only TeV gamma-ray sources are HBL.
So with current instrument sensitivities
there are very strong differences in samples studied
in the different kinds of experiments.
For example, HBL have been followed extensively with the Whipple Observatory 
but have rarely been monitored with EGRET; we therefore
know little about the relative variability at gamma-ray and TeV 
energies (variability within the Compton component) or high-energy 
variability with respect to longer wavelengths, much less the relative
behavior of LBL and HBL at gamma-ray energies.

Our basic picture of blazar continuum emission --- although 
there is still much we do not know --- is that it arises in a fast-moving
jet filled with energetic electrons. Whether the jet is smooth or clumpy 
is not yet clear. If the magnetic field, electron density, and 
particle energy decrease outward along the jet, the highest
energy synchrotron emission comes primarily from the innermost region and 
progressively longer wavelength emission from more extended regions.
The Compton component is presumably produced by scattering of ambient
UV or X-ray photons by the same electrons that are radiating the 
synchrotron photons.
Whether the seed photons are the synchrotron photons themselves
(the synchrotron self-Compton, or SSC, process), or X-ray or UV light from an 
accretion disk, or broad-line photons from the BLR
has yet to be determined. Ghisellini and Madau (1996) explore 
these three options and suggest that the origin may vary from one
kind of blazar to another. In weak-lined blazars like Mrk~421 or 
PKS~2155--304, the SSC model may apply, 
while in strong-lined blazars like 3C~279 (see below), 
the BLR photon density can dominate the local synchrotron photon density 
as seen by the jet electrons (the BLR intensity is enhanced 
due to the bulk relativistic motion of the jet).

There are only a few blazars for which good multiwavelength 
monitoring data are available, and they are very different in luminosity.
Higher luminosity objects (usually RBL) may have physically larger jets
while the HBL are more compact. Below we discuss the variability of 
two of the three best-studied blazars: 3C~279, a strong-emission-line
object (FSRQ), and PKS~2155--304 (HBL).

\section{PRODUCTION OF GAMMA RAYS: THE MULTIWAVELENGTH VARIABILITY OF 3C~279}
\label{sec:3c279}

A key question is, ``How are the gamma-rays produced?'' One
possibility is the BLR photons impinge on the jet, appearing more intense
in the jet frame due to its relative bulk motion, 
and are then scattered by relativistic
electrons in the jet to gamma-ray energies. The BLR clouds nearest the jet
may even be photoionized to a significant extent by the beamed jet 
emission, with essentially zero lag (Ghisellini \& Madau 1996). 
So an important experiment is to
monitor broad-emission-line variability and gamma-ray variability 
simultaneously. It is also critical to monitor the synchrotron light curve
since (a) this reflects the underlying electron population and (b) these
may be the seed photons or may contribute to photoionizing BLR clouds
that then produce the seed photons.

\footnotesize
\begin{figure}
\centerline{\psfig{figure=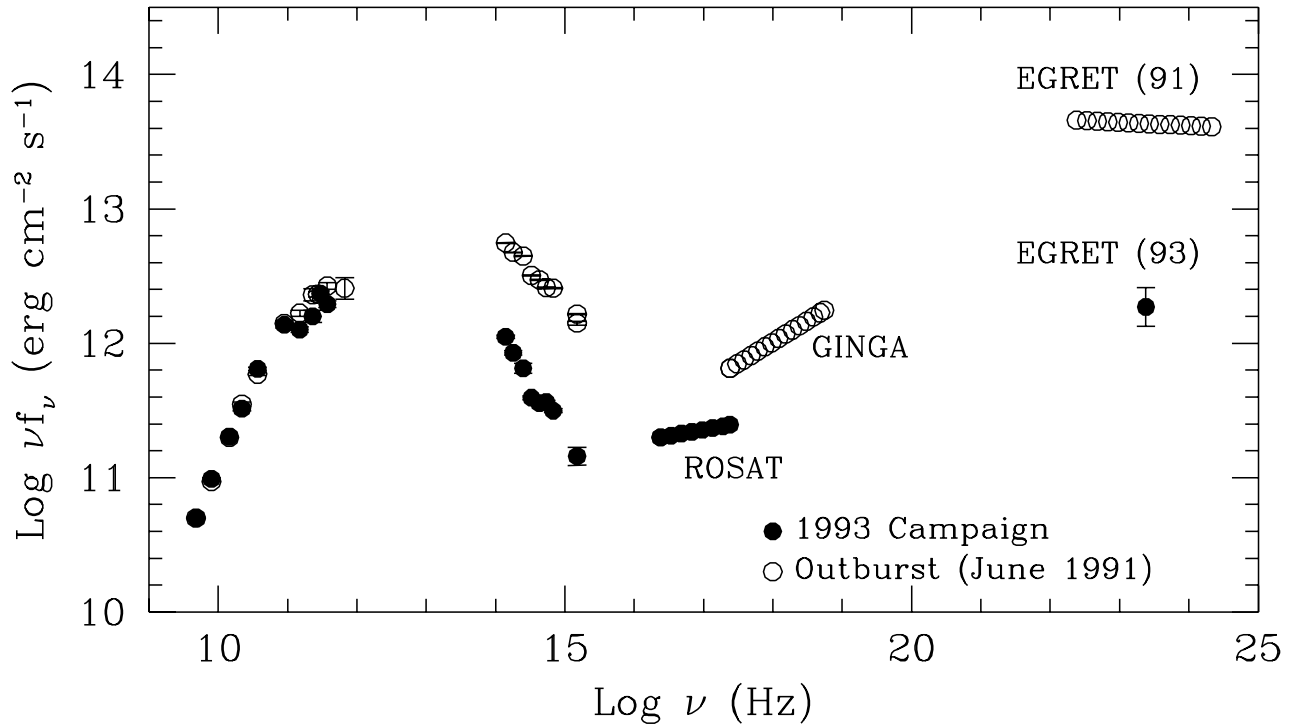,height=3.8in}}


\caption{
Spectral energy distribution of 3C~279 at two epochs. The 
high state was in 1991 June, when 3C~279 was discovered with EGRET,
and the low state was in 1993 January. The variability in the Compton
component (X-gamma-ray energies) is considerably greater than the 
variability in the synchrotron component (IR/optical/UV).
(From Maraschi et al. 1994; copyright American Astronomical Society, 
reproduced with permission.)
}

\end{figure}
\normalsize

Depending on the origin of the blazar variability, different correlations
among wavebands are predicted. First we consider the case where gamma-ray 
variability is caused by variability of the seed photons.
The synchrotron luminosity is proportional to the magnetic 
field strength and to the number density of electrons radiating at the 
particular energy; the inverse-Compton luminosity is proportional to 
the number density of electrons and to the luminosity of the seed photons.
If the seed photons are synchrotron photons (SSC model) 
then the strength of the Compton component is proportional to 
electron density squared (equivalently, to the square of the synchrotron
luminosity). That is, changing only the number density
of energetic electrons creates a much larger variation in gamma-rays
than in the IR/optical/UV.

This same effect can be caused in other ways, however. For example, 
if the seed photons are roughly constant but the jet Doppler factor 
increases slightly, ambient photons impinging on the jet appear brighter 
and the Compton emission grows again by the square of the synchrotron 
luminosity. (The Doppler factor is defined as
$\delta \equiv (\gamma(1-\beta \cos \theta))^{-1}$,
where $\gamma\equiv (1-\beta^2)^{-1/2}$ is the bulk Lorentz 
factor of the jet.)
This is only the case if the seed photons
hit the jet from the front (or side) rather than behind, as might occur
for seed photons from a disk; in that case, an increase in $\delta$ 
would reduce, rather than enhance, 
their apparent intensity to the jet electrons.

If the photoionization of the BLR by the jet is significant,
an increase in either the electrons or the Doppler factor could cause
an intrinsic increase in BLR photons, and thus an even greater increase
in the Compton output.

3C~279 is one of the few blazars for which there are 
extensive radio through gamma-ray data at multiple
epochs. Figure~3 shows the spectrum during a high state in 1991 
and a low state $\sim18$ months later (Maraschi et al. 1994). 
There are two points to note from this figure. First, 
there is much less variability below the synchrotron peak than above it.
This is true over many years of optical and gamma-ray monitoring;
a large flare in early 1996 shows directly that the variability in the 
Compton component is much greater than in the synchrotron component 
(Wehrle et al. 1997).
Second, the relative variations of Compton and synchrotron components
are consistent with the $N^2$ or $\delta^2$ predictions, or with even 
larger relative changes. New data are required to determine which scenario
pertains (the 1996 observations may be sufficient).

\section{JET STRUCTURE: THE MULTIWAVELENGTH VARIABILITY OF PKS 2155--304}

The HBL PKS~2155--304 is the UV-brightest BL Lac and one of the 
X-ray-brightest blazars so it is an obvious target for 
multiwavelength monitoring. There have been three intensive campaigns
on PKS~2155--304, in 1991 November, 1994 May, and 1996 May,
covering optical through X-ray wavelengths. 
(A fourth, predominantly X-ray, campaign occurred in 1996 November.)
Here I discuss results from the first two campaigns and their implications
for blazar jets.

The first campaign consisted of daily observations with IUE and with 
optical and IR telescopes for the full month of 1991 November, 
plus a shorter period
of near-continuous observations with IUE (4.6 days) and Rosat (3.5 days)
in the middle of the month. Only the intensive monitoring resolved the
fast variations; such rapid UV variations had not previously been seen
and were quite unexpected. Flux changes by $\sim10$\% or more
are common on time-scales of half a day.

\footnotesize 
\begin{figure}
\centerline{\psfig{figure=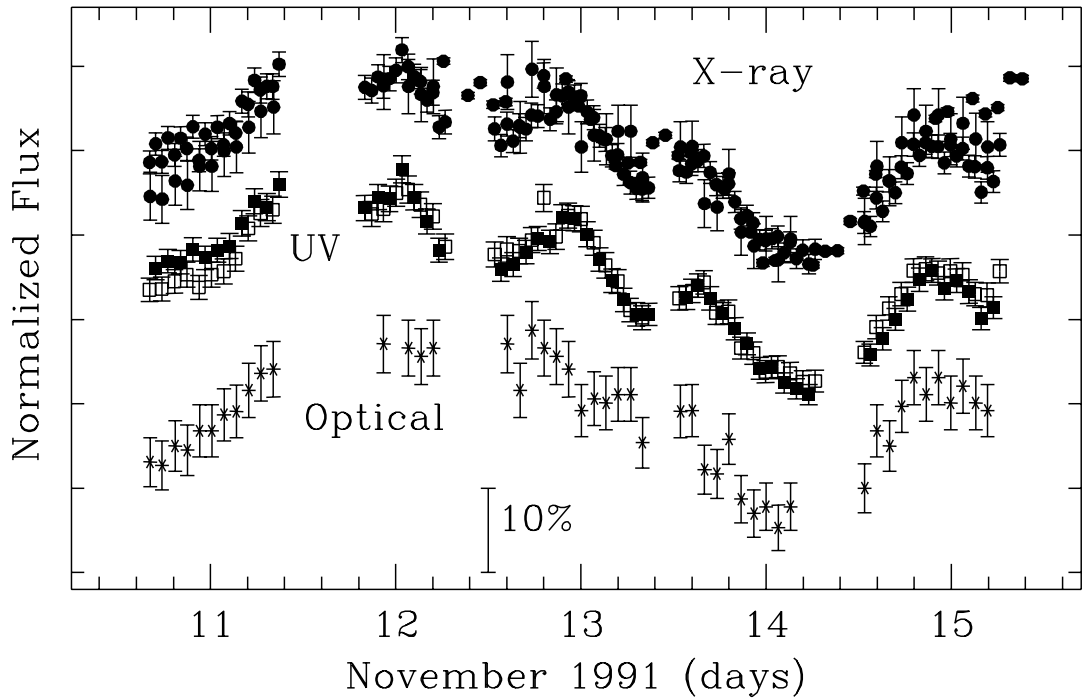,height=3.6in}}


\caption{
Figure 4:  Multiwavelength light curves from intensive monitoring of the
BL Lac object PKS~2155--304 in 1991 November (Edelson et al. 1995).
X-ray data are from the Rosat PSPC; UV data are from the IUE SWP (short
wavelength) and LWP (long-wavelength) spectrographs; optical data are from
the FES monitor on IUE. The emission is closely correlated at all 
wavelengths, and the X-rays lead the UV by $\sim2-3$ hours.
(Copyright American Astronomical Society, 
reproduced with permission.)
\label{fig:2155_1}
}
\end{figure}
\normalsize

The X-ray, UV, and optical light curves from this period, 
shown in Figure~4, are well-correlated, with at most a 2-3-hour lag of UV
photons with respect to X-rays, and their fractional amplitudes are 
independent of wavelength.
(Note that the optical light curve is from data taken with the IUE FES 
[Fine Error Sensor], a modest instrument compared to ground-based systems 
but in this case far superior because of its continuous temporal coverage, 
which is next to impossible from the ground.)

These light curves are intriguing for several reasons. They
established for the first time that X-ray and UV flux were apparently
coming from the same process. This would be expected if the smooth 
UV-X-ray spectrum were produced by the synchrotron process but the wavelength
independence of amplitude was puzzling. In the context of synchrotron 
radiation, higher energy (X-ray-producing) electrons lose energy faster 
than lower energy (UV-producing) electrons and thus cause larger amplitude 
variations within a fixed volume. The relative achromaticity of the 1991
data raises the possibility that another process caused the variations.
Finally, the series of bumps suggest possible recurrent behavior. 

For these reasons, a second campaign was carried out in 1994 May (Fig.~5),
with 12 days of continuous observations of PKS~2155--304 with IUE,
9 days with EUVE (which was integrating long enough to get a good spectrum),
and with ASCA which, because it was newly launched, was on target for only 
2 days.

The ASCA data show a sharp X-ray flare
with amplitude of roughly a factor of 2. The few Rosat observations preceding
the ASCA observation show that similar amplitude X-ray flares must have 
occurred throughout this period. The broad flare in the middle of the IUE 
light curve can plausibly be associated with the X-ray flare 2 days earlier,
particularly given the similar EUV flare which lies between UV and X-ray
both temporally and in wavelength. The amplitude of this flare declines from
$\sim100$\% in the X-ray to $\sim50$\% in the EUV and $\sim35$\% in the UV.
The formal lag of UV with respect to X-ray is 1.7 days, and of UV with
respect to EUV is 1.1 days. Assuming the association of these three
flares is correct, we find striking differences from the 1991 results.
Here the flares are much larger, the lags are much longer
(although the correlation is less definite because of less X-ray coverage),
and the flare amplitude depends strongly on wavelength.

\footnotesize
\begin{figure}
\centerline{\psfig{figure=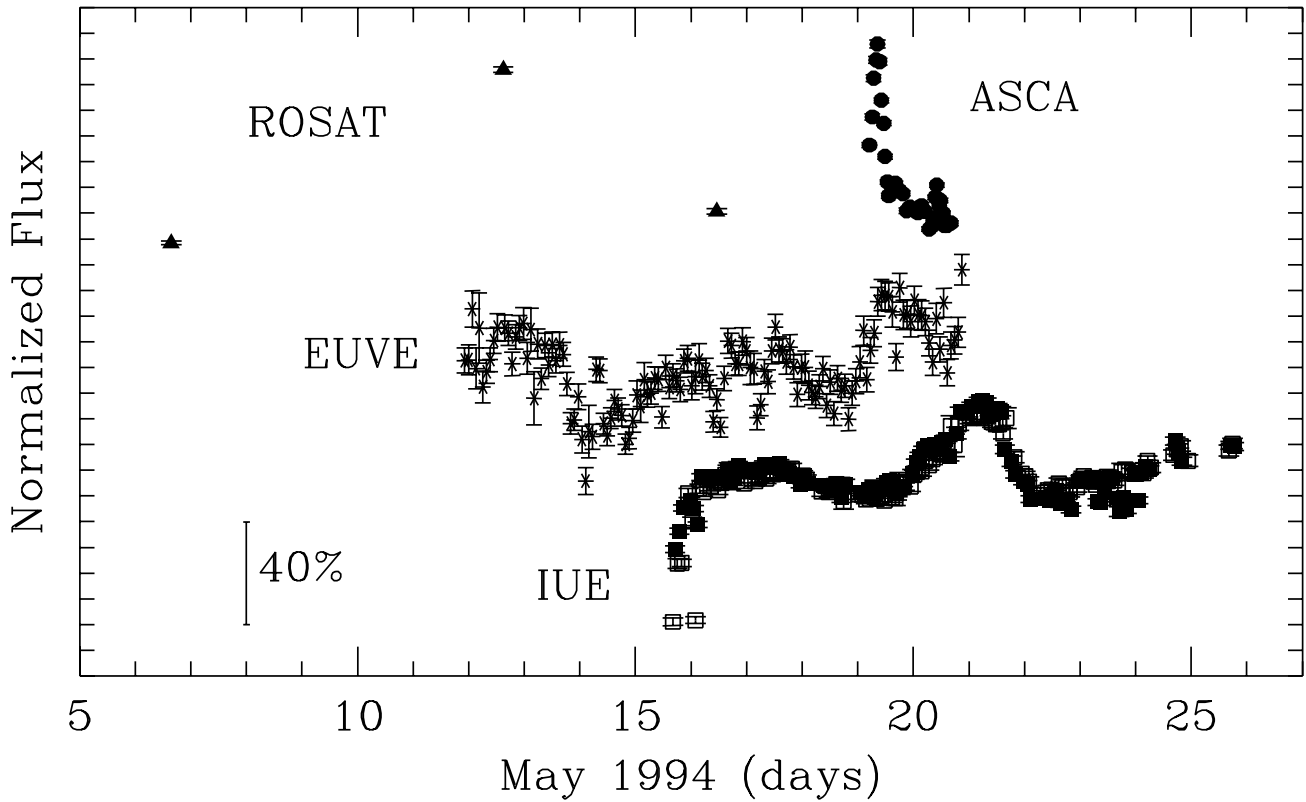,height=4.2in}}


\caption{
Normalized multiwavelength light curves from intensive monitoring 
of the BL Lac object PKS~2155--304 in 1994 May (Urry et al. 1997).
Most of the X-ray data are from ASCA with a few early points from
the Rosat HRI; EUV data are from EUVE;
UV data (with SWP and LWP interspersed) are from IUE. 
The broad flare seen in the middle of IUE 
monitoring seems related to an EUV flare one day earlier and to the
sharp X-ray flare two days earlier. In addition, extremely rapid
(unresolved) variations at the beginning of the IUE observation
have doubling time-scales of 1 hour, faster than previously observed
at UV wavelengths and comparable to the fastest time-scales seen in X-rays.
Compared to 1991, the flares are much larger, the lags are much longer
(although the correlation is less definite because of less X-ray coverage) 
and the flare amplitude declines with increasing wavelength.
\label{fig:2155_2}
}
\end{figure}
\normalsize

In addition, extremely rapid (unresolved) variations at the beginning 
of the IUE observation have doubling time-scales of 1 hour, 
faster than any UV variations previously observed in AGN, and comparable
to the fastest X-ray variations. 
The LWP integration times were half those of the SWP,
which plausibly explains the larger amplitude in the LWP light curve.
Even so, these rapid variations are badly undersampled even with the LWP. 

Figure~6 shows an expanded view of
the beginning of the UV and EUV light curves, with V-band data superimposed.
(Unfortunately, the FES was no longer working well as a photometer for such
faint objects.)
The UV dip is echoed in the V-band polarization; although the latter data
are more poorly sampled than the UV, they start earlier, showing that the
event is indeed a series of dips beginning at the quiescent level. The
sampling is too poor to say whether a similar dip occurs in total
V-band flux, but in general the optical and UV fluxes track each other
well, as was found in the first campaign (Urry et al. 1993).
The EUVE light curve has been shifted horizontally by +1.1 days, which is
the peak of the cross correlation between the UV and EUV light curves 
(most of the power in the cross-correlation comes from the flare several
days later). It does not show as rapid or large-amplitude flaring,
which could indeed have been seen (the EUV bin size is 2/3 the LWP
integration time), but similar structure is apparent in the shifted curve.

The 1994 light curves are clearly different from the
1991 light curves. The central flare is roughly symmetric but broadens 
with wavelength, as the amplitude decreases. Either the physical state of
the emitting region has changed considerably or 
two different mechanisms cause the variations. The latter
possibility is attractive since it is hard to find a single mechanism
that can cause energy-independent variations at one epoch and energy-dependent
variations at another. 
For example, the 1991 fluctuations might be 
caused by microlensing in an intervening galaxy (Treves et al 1997); 
there is strong
Ly$\alpha$ absorption midway to the BL Lac (Bruhweiler et al.
1993) although no galaxy is detected along
the line of sight (van Gorkom et al. 1996).

\footnotesize
\begin{figure}
\centerline{\psfig{figure=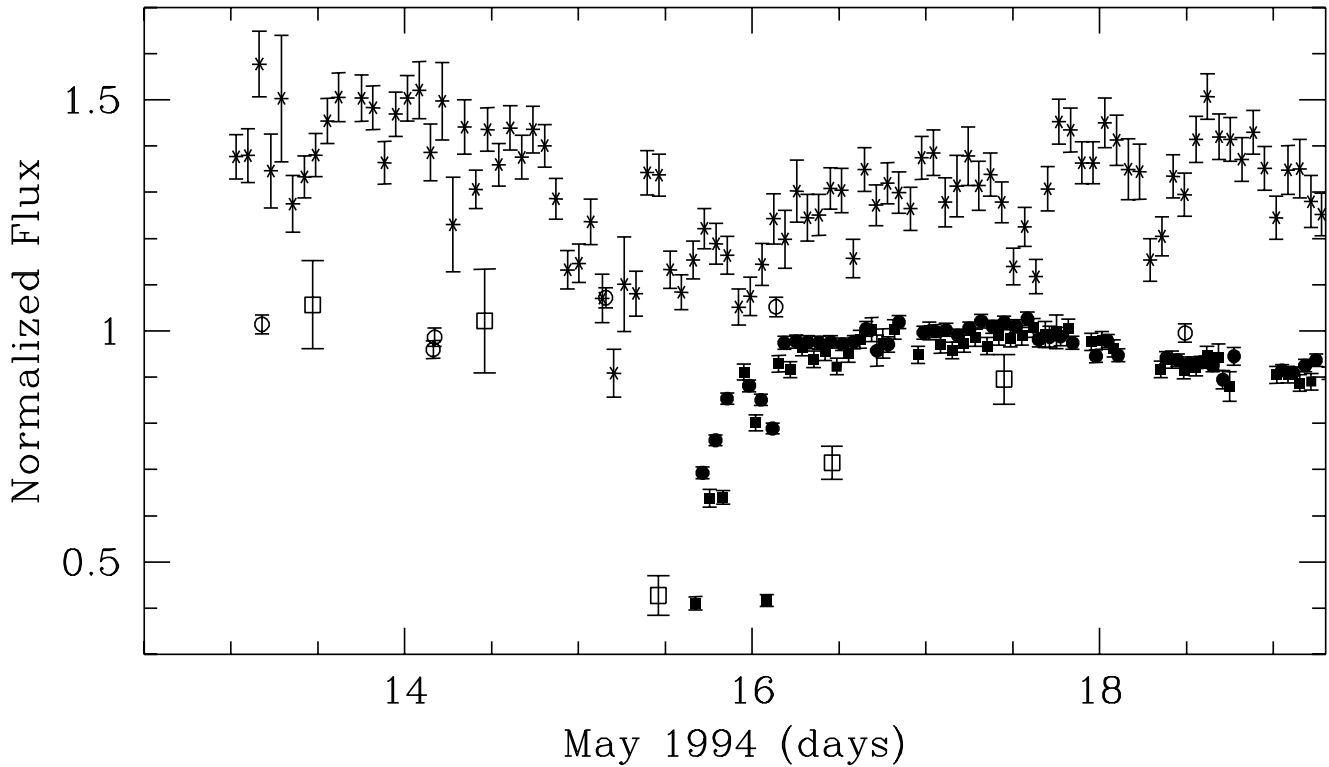,height=4.0in}}


\caption{
Expanded view of the first part of the 1994 monitoring campaign.
{\it Filled points ---} Normalized
UV data for both SWP (circles) and LWP (squares);
{\it stars ---} Normalized EUV data (in 1000~s bins, 
shifted vertically by 0.35 and horizontally by +1.1 days);
{\it open points ---} Normalized V-band (circles) and polarized (squares) 
flux densities.
The UV flux rises by a factor of 2 (in the LWP, for which the integration
time was half that of the SWP), dips again and recovers. 
The polarized V-band flux, though more poorly sampled than the UV, shows
the same dip and recovery.
The shifted EUV light curve shows similar structure though no exact 
corresponding events.
\label{fig:2155_multi}
}
\end{figure}
\normalsize

The central event in the 1994 light curves is consistent with a 
synchrotron flare, in the sense that the variability is of larger 
amplitude and shorter time-scale at higher energies. 
That the plasma is homogeneous can probably be ruled out since in that 
case the flare should begin simultaneously at all wavelengths,
although the decay would be faster at shorter wavelengths. 
This was not observed.
The delays between wavebands are comparable to the time-scales of the flare, 
as expected if a disturbance (e.g., shock or compression wave)
were propagating through an inhomogeneous emitting region. 
Such a region could arise naturally as magnetic fields and electron
densities vary globally or in the steady-state situation behind a 
standing shock. However, the decay times appear to be slightly longer 
than the lag times, which is difficult to understand if the lags 
and/or the spatial stratification are dominated by the radiative time-scale.
For further discussion, see Urry et al. (1997).

\section{WHAT WE NOW KNOW ABOUT BLAZARS}

Current observations suggest that the light from blazars is dominated 
by emission from a relativistic jet closely aligned with the line of sight.
The broad-band spectrum of this light is consistent with two separate
components. At long wavelengths, a synchrotron component peaks 
(in $\nu F_\nu$) in the IR-optical regime for FSRQ and LBL and in the 
UV-X-ray regime for HBL. At short wavelengths, a Compton component peaks
at GeV-TeV energies or above; the shape and variability
of this component are much more poorly constrained because of sparse data.
The strong gamma-ray emission confirms that the observed radiation must
be relativistically beamed or else the pair-production opacity would be
too high for gamma-rays to escape.

The same electrons are likely responsible for the emission in both 
synchrotron and Compton components. 
The seed photons scattered to gamma-ray energies may be
synchrotron photons for weak-lined blazars (BL Lacs) and UV
photons from the BLR in strong-lined blazars. This can be tested by
direct observation of the correlated variations in these components.

To date, 3C~279 and PKS~2155--304 are two of the three best-monitored
blazars (the third is Mrk~421). Available data suggest that the
synchrotron emission comes from an inhomogeneous region. Further data
are needed to determine the physical state of the jet, particularly
well-sampled light curves at the highest energies. The ASCA light curves
of PKS~2155--304 and Mrk~421, for which the X-ray emission is part of
the synchrotron component, both show the hard X-rays leading the soft
X-rays by $\sim1$~hour (Makino et al. 1996, Takahashi et al. 1996).
The frequency dependence of the lag goes roughly as $\nu^{-1/2}$ (this
holds approximately for the EUV and UV as well in PKS~2155--304), as
expected if the delays are related to the synchrotron cooling time.

Finally, blazars exhibit no obvious signatures of accretion disks: they
do not have big blue bumps and indeed the simultaneity of 
variations in the optical and UV continuum emission rules out an
origin in a standard accretion disk (Urry et al. 1993).
On the other hand, that jets are present provides a theoretical
argument for disks, since they may be important for jet formation. 
At the same time, there may be evidence for winds, possibly of material
evaporating from an accretion disk. 
Strong absorption from highly ionized gas has been seen in the EUV and X-ray
spectra of a few BL Lacertae objects (Canizares \& Kruper 1984, Madejski
et al. 1991, K\"onigl et al. 1995). Interpreting this absorption as a wind 
implies a very massive, very high velocity outflow of highly ionized matter,
either in the jet or close to the central engine.

\acknow
This paper includes results from a large, ongoing, collaborative effort to
monitor blazars at multiple wavelengths. I thank the many participants in 
these campaigns for helpful discussions, especially
Laura Maraschi, Joe Pesce, 
Elena Pian, Rita Sambruna, Aldo Treves, and Ann Wehrle. 
Helpful comments were also provided by an anonymous referee.
I am grateful to Joe Pesce, Elena
Pian, John Godfrey, and Harry Payne for help in preparing the manuscript.
This work was supported in part by NASA grants NAG5-2510, 
NAG5-2538, and NAG5-3138.

\end{document}